\documentclass{article}
\usepackage{geometry}
\usepackage{color}
\usepackage{spconf,amsmath,graphicx}
\usepackage{bm}
\usepackage{amsthm}
\usepackage{assoccnt}
\usepackage{amsmath,subcaption}
\usepackage{cite}
\usepackage{psfrag}
\usepackage[cmintegrals]{newtxmath}
\def\x{{\bm{x}}}
\def\xk{{\bm{x}}(k)}
\def\xkk{{\bm{x}}(k+1)}
\def\y{{\bm{y}}}
\def\yk{{\bm{y}}(k)}
\def\A{{\bm{A}}}

\def\B{{\bm{B}}}
\def\u{{\bm{u}}}
\def\uk{{\bm{u}}(k)}
\def\D{{\bm{D}}}
\def\C{{\bm{C}}}
\def\P{{\mathcal{P}}}
\def\G{{\mathcal{G}}}
\def\V{{\mathcal{V}}}
\def\L{\bm S}
\def\Ts{\mathcal{T}_s}

\def\E{{\mathcal{E}}}

\def\Rr{{\mathbb{R}}}
\def\Cn{{\mathcal{C}_{n}}}
\def\On{{\mathcal{O}_{n}}}
\newcommand{\rnk}[1]{{\rm rank}(#1)}

\def\U{{\bm{U}}}

\newtheorem{mydef}{Definition}
\newtheorem{mylemma}{Lemma}

\title{State-Space Based Network Topology Identification}
\name{Mario Coutino$^{\dagger}$, Elvin Isufi$^\dagger$, Takanori Maehara$^\ddagger$, Geert Leus$^\dagger$\thanks{This research is supported in part by the ASPIRE project (project 14926 within the STW OTP programme), financed by the Netherlands Organization for Scientific Research (NWO). Mario Coutino is partially supported by CONACYT and AIP RIKEN.}}
\address{Delft University of Technology, Delft, The Netherlands$^\dagger$\\ AIP RIKEN, Tokyo, Japan$^\ddagger$}

\begin{document}
\ninept
\maketitle
\captionsetup[figure]{font=scriptsize,labelfont=scriptsize}
\begin{abstract}
In this work, we explore the state-space formulation of network processes to recover the underlying structure of the network (local connections). To do so, we employ subspace techniques borrowed from system identification literature and extend them to the network topology inference problem. This approach provides a unified view of the traditional network control theory and signal processing on networks. In addition, it provides theoretical guarantees for the recovery of the topological structure of a deterministic linear dynamical system from input-output observations even though the input and state evolution networks can be different.
\end{abstract}

\section{Introduction} \label{sec.intro}

In recent years, major efforts have been focused on extend traditional tools from signal processing to cases where the acquired data is not defined over typical domains such as time or space but over a network (graph)~\cite{kolaczyk2009statistical,8347160}. The main reason for the increase of research in this area is due to the fact that network-supported signals can model complex processes. For example, by means of signals supported on graphs we are able to model transportation networks~\cite{deri2016new}, brain activity~\cite{sporns2010networks}, and epidemic diffusions or gene regulatory networks~\cite{mittler2004reactive}, to name a few.

As modern signal processing techniques take into account the network structure to provide signal estimators~\cite{narang2013signal,di2017adaptive,chen2015signalrecovery}, filters~\cite{Coutino2017,shuman2013emerging,isufi2017autoregressive,coutino2018advances}, or detectors~\cite{hu2016matched,chepuri2016subgraph,isufi2018Blind}, appropriate knowledge of the interconnections of the network is required. In many instances, the knowledge of the network structure is given and can be used to enhance traditional signal processing algorithms. However, in other cases, the network information is unknown and needs to be estimated. As the importance of studying such structures in the data has been noticed, retrieving the topology of the network has become a topic of extensive research~\cite{kalofolias2016learn,mateos2018inference,segarra2017network,karanikolas2016multi,shen2017kernel,lu2017closed,zhou2007topology,shen2017tensor,shafipour2017network}. 

Despite the extensive research done so far (for a comprehensive review the reader is referred to~\cite{mateos2018inference,8347160} and references therein), most of the approaches do not lever a physical model beyond the one induced by the so-called \emph{graph filters}~\cite{marques2017stationary,segarra2017network} drawn from graph signal processing (GSP)~\cite{shuman2013emerging,sandryhaila2014big,ortega2018graph}. Among the ones that propose a different interaction model e.g.,~\cite{shafipour2017network,shen2017kernel}, none of them considers the network data (a.k.a. graph signals) as states of an underlying process nor considers the that the input and the state may evolve on different underlying networks. However, different physical systems of practical interest can be defined through a state-space formulation with (probably) known functions, i.e., brain activity diffusion, finite element models, circuit/flow systems. For these processes, a more general approach to find the underlying connections is required. In this work, we therefore focus on the general problem of \emph{retrieving the underlying network structure}, from input-output signals, of a process that can be modeled through a \emph{deterministic linear dynamical system} whose system matrices depend on the interconnection of the network. 
\section{State-Space Models for Network Processes} \label{sec.model}
Let us consider a tuple of graphs $\G_1 = \{\V,\E_1\}$ and $\G_2 = \{\V,\E_2\}$ to represent two networks, where $\V=\{v_{1},\ldots,v_{n}\}$ and $\E_i  \subseteq \V\times\V$ for $i \in \{1,2\}$ denote their vertex and edge sets, respectively. Further, let $\P$ be a process over $\{\G_1,\G_2\}$ that describes the evolution through time of a signal (the state) $\x(t)$ defined over $\G_1$ coupled with another signal (the input signal) defined over $\G_2$. Such process can be described through the linear dynamical system
\begin{subequations}\label{eq.ltia}
	\begin{eqnarray}
		\xkk &=& f_1(\L_1)\xk + f_2(\L_2)\uk\;\in \Rr^{n},\\
		\yk &=& \C\xk + \D\uk\;\in\Rr^{l},
	\end{eqnarray}
\end{subequations}
where $\L_i,\,i\in\{1,2\},$ is the matrix representation of the graph $\G_i$, i.e., the shift operator in the GSP terminology, $\C\in\mathbb{R}^{l\times n}$, and $\D\in\mathbb{R}^{l\times n}$ are the observation matrices and $f_i : \mathbb{R}^{N\times N} \rightarrow \mathbb{R}^{N\times N}$ is a matrix function defined via the Cauchy integral by~\cite{higham2008functions}
\begin{equation}\label{eq.intDef}
	f_i(\L) := \frac{1}{2\pi i}\int_{\Gamma_{f_i}}f_i^{\rm s}(z)R(z,\L)dz, 
\end{equation}
with $f_i^{\rm s}$ being the scalar version of $f_i$ which is assumed analytic on and over the contour $\Gamma_{f_i}$. Here, $R(z,\L)$ is the resolvent of $\L$ given by
\begin{equation}
	R(z,\L) := (\L - z\bm I)^{-1}.
\end{equation}

Model \eqref{eq.ltia} is expressed in terms of its \emph{state-space} representation and captures the relation between the input, the output, and the state through a first-order difference equation~\cite{ogata2002modern}. It connects the output (observables), $\yk$, to a set of variables (states), $\xk$, which vary over time and depend on their previous value and on external inputs (excitations), $\uk$.

After observing model~\eqref{eq.ltia}, a natural question that arises is the following: \emph{assuming that the observation matrix $\C$ and the relation between $\P$ and $\{\G_i,\G_j\}$ are known, how can we retrieve $\{\L_i,\L_j\}$, i.e., the network structures, from a number of samples of the input signal $\uk$ and the output signal $\yk$?} 

In this work, we aim to answer this question by employing techniques commonly used in control theory which rely on results for Hankel matrices and linear algebra. In particular, we employ subspace techniques which do not require any parametrization of the model, hence the problem of performing nonlinear optimization, as in the prediction-error methods~\cite{ljung2002prediction}, is avoided.

\section{Identifiability Conditions for LTI Systems}
For the sake of simplifying notation, from this point on, we omit the dependency on $\L_i$ and $\L_j$ of the system matrices in~\eqref{eq.ltia} and refer to the matrices $f_i(\L_i)$ and $f_j(\L_j)$ as $\A$ and $\B$, respectively.

Prior to introducing the methods for network topology identification, we digress the identifiability conditions of LTI systems. In this section, we briefly recap the requirements on the system matrices $(\A,\B,\C,\D)$ for applying subspace techniques for estimating them.

The main requirement for system identification is the \emph{minimality} condition of system~\eqref{eq.ltia}. This property is intrinsically related to two well-known properties of dynamical systems: \emph{reachability} and \emph{observability}. The first property denotes the ability of the input, $\uk$, to steer the system state to the zero state within a finite time interval. While the second denotes the ability to observe the time evolution of the states through the evolution of the output; that is, it answers the question of the uniqueness of the relation between state and the output.

Before stating these notions mathematically, let us introduce the following two matrices~\cite{ogata2002modern}
\begin{itemize}
	\item \textbf{Controllability Matrix: } 
		$\mathcal{C}_s \triangleq [\B\;\A\B\;\cdots\;\A^{s-1}\B],$
	\item \textbf{Observability Matrix: } 
		$\mathcal{O}_s \triangleq [\C^T\;\cdots\;(\A^{s-1})^T\C^T]^T.$
\end{itemize}
Based on these matrices, the following two lemmas state the concepts of reachability and observability in a more formal way.
\begin{mylemma}\textnormal{(Reachability)}
The LTI system~\eqref{eq.ltia} is reachable if and only if $\rnk{\Cn} = n$.
\end{mylemma}

\begin{mylemma}\textnormal{(Observability)}
The LTI system~\eqref{eq.ltia} is observable if and only if $\rnk{\On} = n$.
\end{mylemma}
By using these results we can now formally state the definition of \emph{minimality} of a system.
\begin{mydef}\textnormal{(Minimality)}
The LTI system~\eqref{eq.ltia} is minimal if and only if it is both reachable and observable. Furthermore, the dimension of the state vector $\xk$ of the minimal system defines the \emph{order} of the LTI system.
\end{mydef}

As the system identification framework only guarantees recovery of a \emph{minimal} system, from this point on, we only consider problem instances where the system of interest is minimal. Note that this is not a restrictive assumption, as even when we retrieve a minimal system of order $p<n$, this can be interpreted as a system on the nodes of a hypergraph, i.e., clusters of nodes that drive the general behavior of the process over the network.

\section{Subspace Network Identification} \label{sec.netid}

In this section, we introduce a general framework for estimating the topology of the networks, i.e., the associated matrices $\{\L_1,\L_2\}$, from input-output relations. To do so, we first provide the methods for retrieving the system matrices in \eqref{eq.ltia}. Then, we state the required conditions and propose different methods for estimating the graph matrices $\{\L_1,\L_2\}$ from the obtained system matrices.

\subsection{State-Space Identification}

It is not hard to show that the state of the system~\eqref{eq.ltia} with initial state $\bm x(0)$ at time instant $k$ is given by
\begin{equation}\label{eq.xk}
	\xk = \A^k\x(0) + \sum\limits_{i=0}^{k-1}\A^{k-i-1}\B\u(i).
\end{equation}

Observing the expression relating the states, the input and the output in~\eqref{eq.ltia}, we can specify the following relationship between the batch input $\{\bm u(k)\}_{k=0}^{s-1}$ and the batch output $\{\bm y(k)\}_{k=0}^{s-1}$
\begin{equation}
	\begin{bmatrix}
		\bm y(0)\\
		\vdots\\
		\bm y(s-1)
	\end{bmatrix} = \mathcal{O}_s\x(0) + \Ts\begin{bmatrix}
	\u(0)\\
	\vdots\\
	\u(s-1)
	\end{bmatrix},
\end{equation}
where 
\begin{equation}
	\Ts \triangleq \begin{bmatrix}
	\D & 	\bm 0 	& \bm 0 & \cdots & \bm 0 \\
	\C\B & 	\D 		& \bm 0	& \cdots & \bm 0 \\
	\C\A\B & \C\B & \D  & \cdots & \bm 0 \\
	\ldots & & \ddots & \ddots &  \\
	\C\A^{s-2}\B 	& \C\A^{s-3}\B & \ldots & \C\B & \D
	\end{bmatrix},
\end{equation}
and $s$ is the size of the batch that must be larger than the number of states (assuming the number of nodes is the number of states this implies $s > n$).

Given that the underlying system is time-invariant (i.e., the graph does not change in time), the following relation holds~\cite{verhaegen2007filtering} 
\begin{equation}\label{eq.dataEq}
	\bm Y_{m} = \mathcal{O}_{s}\bm X_{m} + \Ts\bm U_{m}
\end{equation}
where
$\bm X_{m} \triangleq [\x(0),\x(1),\cdots,\x(m-1)]$, $\bm y_{i,s} \triangleq [\y(i)^T,\y(i+1)^T,\cdots,\y(i+s-1)^T]^T$, $\bm Y_{m} \triangleq [\bm y_{0,s},\bm y_{1,s}, \cdots,\bm y_{m-1,s}]$, $\bm u_{i,s} \triangleq [\u(i)^T,\u(i+1)^T,\cdots,\u(i+s-1)^T]^T$ and $\bm U_{m} \triangleq [\bm u_{0,s},\bm u_{1,s},\cdots,\bm u_{m-1,s}]$ with $m > s$.

Throughout this work, we assume that that $\C$ has rank equal to $n$. Despite that this assumption seems restrictive, we consider it to simplify the exposition of the approach. Dealing with dynamical models whose output dynamics satisfy $l < n$ is not trivial. As it will become evident, disambiguation of the system matrices requires extra information when $\C$ is wide or singular. Therefore, this is left for immediate future work.

To identify the system matrices from~\eqref{eq.ltia}, we first make use of the following lemma.
\begin{mylemma}(Verhaegen and Dewilde~\cite{verahegen1992subspace}) Given the following RQ factorization
	\begin{equation}
		\begin{bmatrix}
			\bm U_{m}\\ \bm Y_{m} 
		\end{bmatrix} = \begin{bmatrix}
			\bm R_{11} & \bm 0 & \bm 0 \\ \bm R_{21} & \bm R_{22} & \bm 0 
		\end{bmatrix}\begin{bmatrix}
			\bm Q_1 \\ \bm Q_2 \\ \bm Q_3
		\end{bmatrix},
	\end{equation}
	for appropriately sized matrices $\bm R$ and $\bm Q$, the following relationship holds for the input-output data matrices
	\begin{equation}
		\bm Y_{m}\bm\Pi_{m}^{\perp} = \bm R_{22}\bm Q_2,\;\; \text{where }~\bm U_{m}\bm\Pi_{m}^{\perp} = \bm 0.
	\end{equation}
	\label{lm.rq}
\end{mylemma}
Using Lemma~\ref{lm.rq}, it can be shown that 
\begin{equation}
{\rm range}(\bm Y_{m}\bm\Pi_{m}^{\perp}) = {\rm range}(\mathcal{O}_s) = {\rm range}(\bm R_{22}\bm Q_2).
\end{equation}
Therefore, from the singular value decomposition (SVD) of $\bm R_{22}$, i.e., $\bm R_{22} = \bm U_R\bm \Sigma_R \bm V_R^T$, we can obtain the transition matrix $\A$ (up to a similarity transform) as follows. First, from
\begin{equation}\label{eq.Ur}
	\bm U_R = \mathcal{O}_s\bm T = \begin{bmatrix}
	\C \bm T\\ \C\bm T(\bm T^{-1}\bm A\bm T)\\ \vdots \\ \C\bm T(\bm T^{-1}\bm A\bm T)^{s-1}
	\end{bmatrix} = \begin{bmatrix}
		\C_T \\ \C_T\A_T \\ \vdots \\ \C_T\A_T^{s-1}
	\end{bmatrix},
\end{equation}
where we have defined $\A_T := \bm T^{-1}\A\bm T$ and $\C_T := \bm C\bm T$ for an unknown similarity transformation matrix $\bm T\in\Rr^{n\times n}$, we can compute an estimate $\hat{\A}_T$ of  $\A_T$ by solving the overdetermined system 
\begin{equation}\label{eq.systAt}
	\bm U_{R,l}\bm A_T = \bm U_{R,r},
\end{equation}
which exploits the shift-invariance of the system. Here, we have defined the matrices
\begin{eqnarray}
     \bm U_{R,r} &:=& \bm U_R(l+1:sl,:),\\
     \bm U_{R,l} &:=& \bm U_{R}(1:(s-1)l,:),
\end{eqnarray}
and abused the MATLAB notation to denote the rows and columns that are considered for building system~\eqref{eq.systAt}. From~\eqref{eq.Ur}, we can observe that an estimate $\hat\C_T$ of $\C_T$ can be obtained by selecting the first $l$ rows of $\bm U_R$.

Since of $\C$ is full rank, we can estimate the similarity transform $\bm T$ from $\C_T$. Therefore, the estimate $\hat{\A}$, for $\A$, can be obtained as
\begin{equation}\label{eq.ahat}
	\hat{\A} = \C^{-1}\hat{\C}_T\hat{\A}_T(\hat{\C}_T)^{-1}\C.
\end{equation}

While a similar approach using the matrices $\bm R_{21}$ and $\bm R_{11}$ can be performed for retrieving a transformed $\B$, i.e.,  $\B_{T} = \bm T^{-1}\B$, ~\cite{verahegen1992subspace}, here we compute it, together with the initial state $\x_T(0) = \bm T^{-1}\x(0)$, by solving a least squares problem. This is done to keep the exposition of the approach conceptually simple, as the usage of the information in $\bm R_{21}$ and $\bm R_{11}$ requires the introduction of another (more involved) shift-invariant structure.

To do so, first, observe that for given matrices $\A_T$ and $\C_T$ the output can be expressed linearly in the matrices $\B_T$ and $\D$ as
\begin{equation}
	\begin{aligned}
	\yk{}&= \C_T\A_T^k\x_T(0) + \\ &\bigg(\sum\limits_{q=0}^{k-1}\bm u(q)^T\otimes \C_T\A_T^{k-q-1}\bigg){\rm vec}(\B_T) + (\bm u(k)^T \otimes \bm I_l){\rm vec}(\D),
	\end{aligned}
\end{equation}
where $\bm I_l$ is the $l\times l$-identity matrix. From here, by defining $\bm\theta = [\bm x_{T}(0)^T\;{\rm vec}(\bm B_T)^T\;{\rm vec}(\bm D)^T]^T$, we can find the system matrices by solving
\begin{equation}
	\begin{array}{ll}
		\underset{\bm\theta}{\min}\frac{1}{m}\sum_{k=0}^{m-1}\Vert \y(k) - \bm\Psi\bm\theta\Vert_2^2,
	\end{array}
\end{equation}
where $\bm\Psi \triangleq [\C_T\A_T^{k},\; (\sum\limits_{q=0}^{k-1}\bm u(q)^T\otimes \C_T\A_T^{k-q-1}),\; (\bm u(k)^T\otimes\bm I_l)]$.

After the estimates $\hat{\B}_T$ and $\hat{\D}$ are obtained, we can solve for the original matrices by appropriately multiplying them with the estimate of the similarity transform as we did to retrieve $\A$ [cf.~\eqref{eq.ahat}].
\setcounter{figure}{1}
\begin{figure*}[ht!]
    \centering
    \psfrag{True}{\scriptsize{True}}
    \psfrag{Estimated}{\scriptsize{Estimated}}
    \psfrag{Network of States}{}
    \psfrag{nz = 30}{}
    \begin{subfigure}{0.3\textwidth}
        \includegraphics[width = \textwidth]{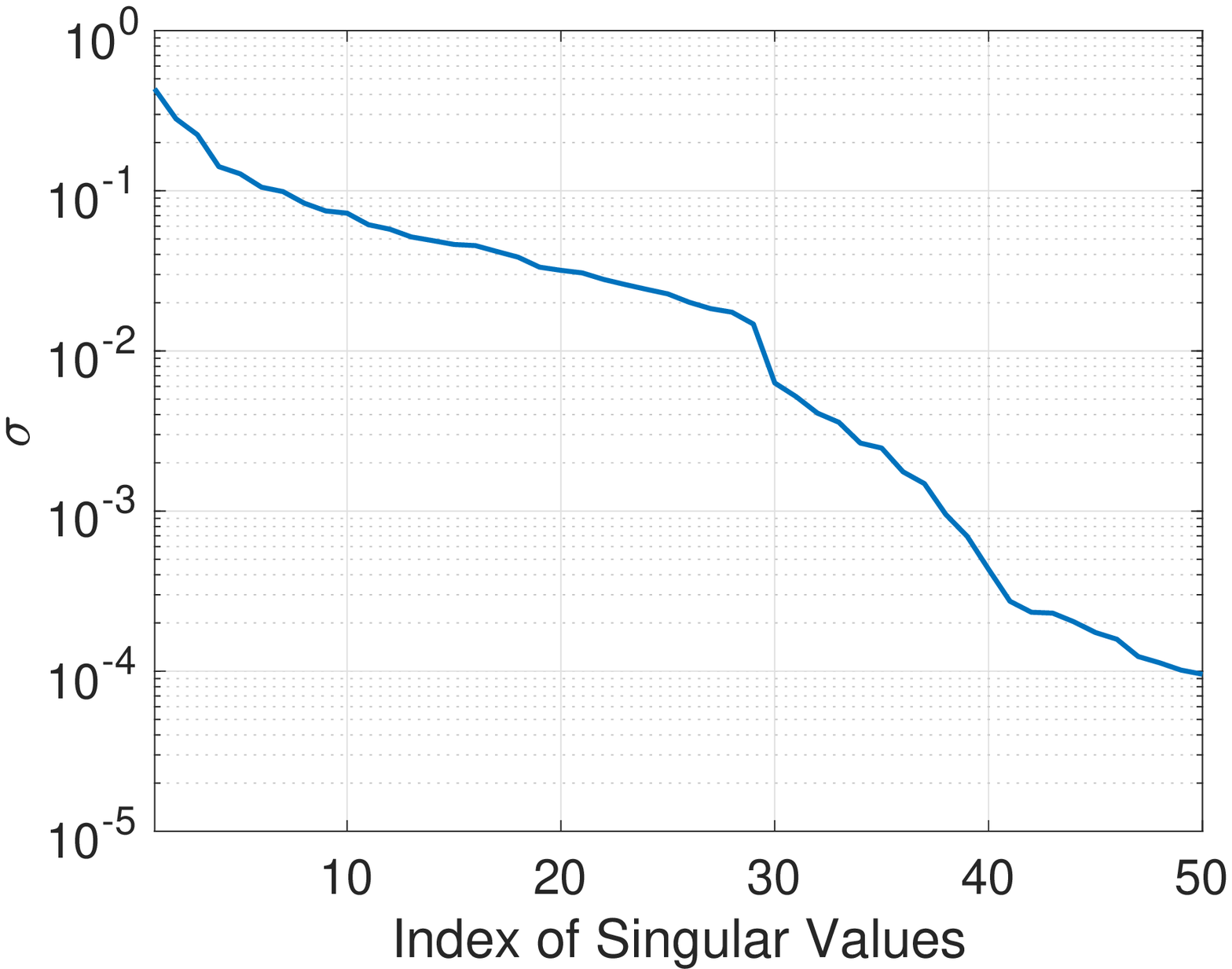}
        \caption{}
        \label{fig.g1}
    \end{subfigure}%
    \begin{subfigure}{0.3\textwidth}
        \psfrag{Network of Input}{}
        \includegraphics[width = \textwidth]{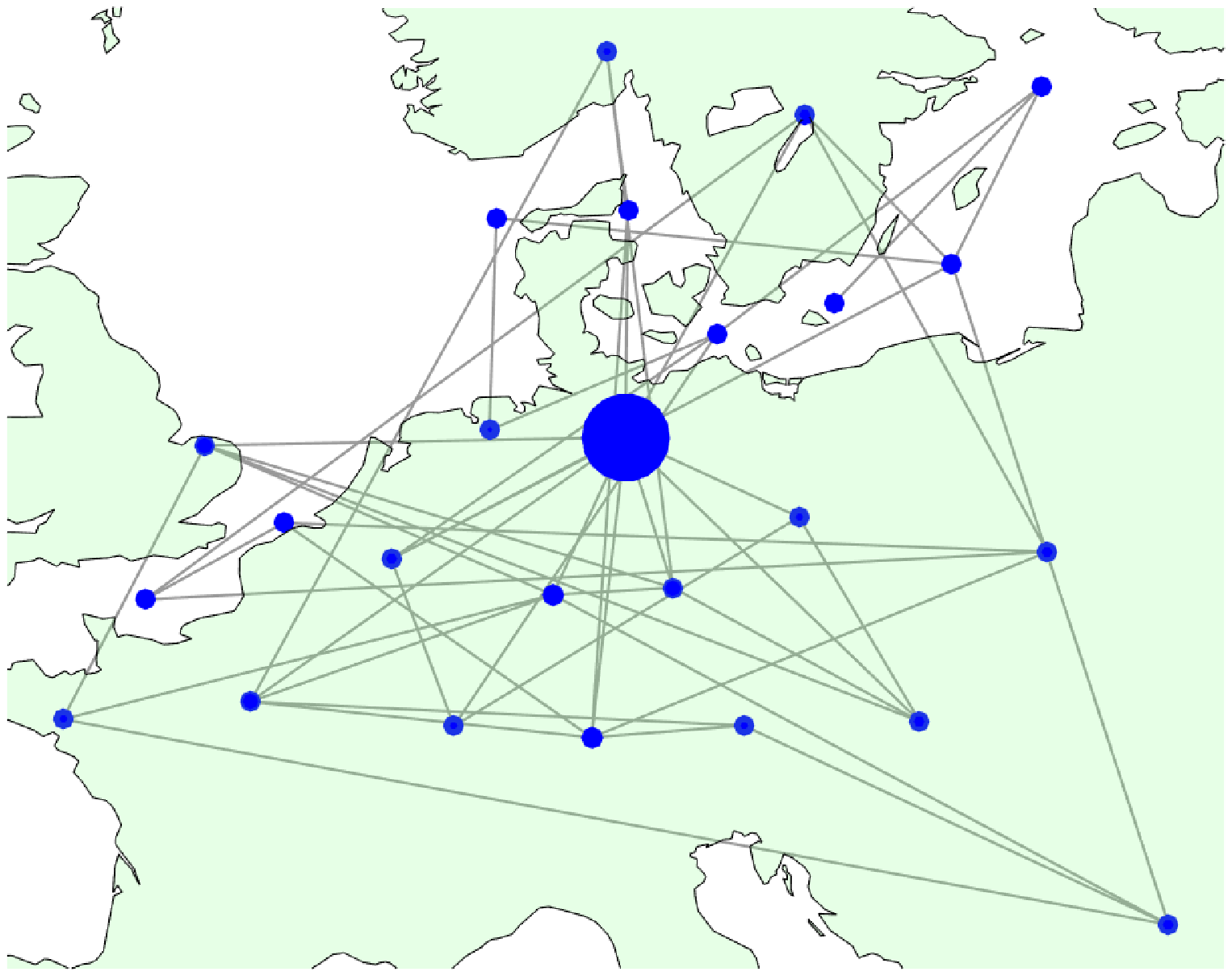}
        \caption{}
        \label{fig.g2}
    \end{subfigure}%
    \begin{subfigure}{0.3\textwidth}
        \includegraphics[width = 0.9\textwidth]{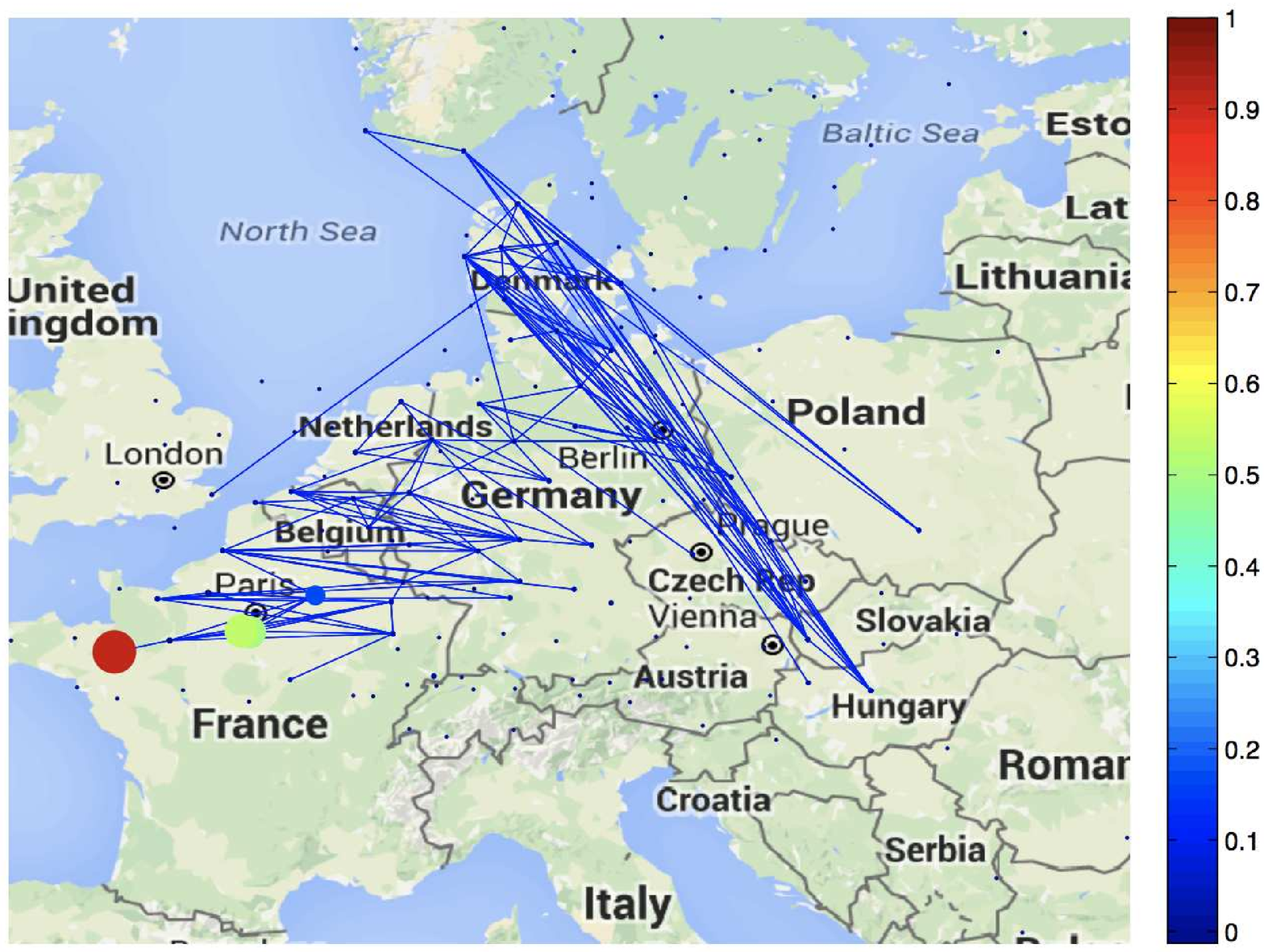}
        \caption{}
        \label{fig.g3}
    \end{subfigure}
    \caption{(a) First $50$ singular values ($\sigma$) of $\frac{1}{m}\bm Y_m$ for $m =3\times10^4$ and $s = 169$. Here, the original dataset has been interpolated and filter to increase the length of the recordings. The first knee, i.e., the drop in singular values (SV), occurs at the $5$th SV, the second drop happens around the $20$th SV. (b) Learned graph from dataset using the proposed one-shot state graph estimator. (c) Learned graph in~\cite{thanou2017learning} using the ETEX dataset.}
    \label{fig.fullG}
\end{figure*}
\subsection{Network Identification}
At this point, the system matrices have been obtained. Now, we consider different scenarios for estimating the topology of the underlying networks.
 
\textbf{Known Scalar Mappings $\{f_1^{\rm s},f_2^{\rm s}\}$.} In this case, we first obtain the eigenvalues of the graph matrices by applying the inverse mappings $(f_{1}^{\rm s})^{-1}$ and $(f_{2}^{\rm s})^{-1}$ to the spectra of the respective matrices. Therefore, for guaranteeing a unique set of eigenvalues for the graph matrices, the functions $\{f_{1}^{\rm s},f_2^{\rm s}\}$ should be bijective, i.e., a one-to-one mapping, on an appropriate domain. For instance, for $\L_i$ being the normalized Laplacian, the mappings should be bijective in the interval $[0,2]$ as the spectrum of the normalized Laplacian lies there.

When the inverse mappings cannot be found analytically (e.g., due to computational reasons), the problem of finding the eigenvalues of the graph matrices boils down to a series of \emph{root finding} problems. That is, consider $[\bm\omega_i]_{k}$ as the $k$th eigenvalue of the matrix $\bm M_i$, where $\bm M_1 := \hat{\A}$ and $\bm M_2 := \hat{\B}$, and $f_i^s$ is the known scalar mapping. Then, the estimation of the eigenvalue vector $\bm\lambda_i$ for each of the matrices can be formulated as
\begin{equation}
    \begin{array}{ll}
        [\hat{\bm\lambda}_i]_k = \underset{[\bm \lambda_i]_k\in\Rr^+}{\text{arg min}} & \Vert f_i^{\rm s}([\bm \lambda_i]_k) - [\bm\omega_i]_k \Vert_2^2, 
    \end{array}
\end{equation}
for $i \in \{1,2\}$. Fortunately, there exist efficient algorithms to obtain roots with a high accuracy even for non-linear functions~\cite{stoer2013introduction}. In addition, note that even when only $\bm A_T$ is known, we can still retrieve the eigenvalues of $\L_i$ as this matrix is \emph{similar} to $\A$, i.e., $\A_T = \bm T^{-1}\A\bm T$  . As by definition $\hat{\A}$ and $\hat{\B}$ are matrix functions of $\L_1$ and $\L_2$ [cf.~\eqref{eq.intDef}], respectively, we can use the eigenbasis from these matrices to reconstruct the graph matrices as
\begin{equation}
    \hat{\L}_i = \hat\U_{i}{\rm diag}(\hat{\bm\lambda}_i)\hat\U_i^{-1},\;\text{for }i\in\{1,2\},
\end{equation}
with $\hat\U_1 = {\rm eigvecs}(\hat{\A})$ and $\hat\U_2 = {\rm eigvecs}(\hat\B)$.

\textbf{Unknown Scalar Mappings.} When the scalar functions $\{f_1^{\rm s},f_2^{\rm s}\}$ are unknown, we can opt to retrieve the \emph{sparsest} graphs that are able to generate the estimated matrices, i.e.,
\begin{equation}\label{eq.Lnorm0}
	\begin{array}{lll}
		\hat{\L}_i = & \underset{\bm\omega_i\in\mathbb{R}^{n}}{\rm argmin} & \Vert \L_i \Vert_0 \\
		& \text{subject to} & \L_i = \hat{\U}_i{\rm diag}(\bm\omega_i)\hat{\U}_i^{-1}, \;\;\L \in \mathcal{S},
	\end{array}
\end{equation}
where ${\rm diag}(\cdot)$ denotes a diagonal matrix with its argument on the main diagonal and $\mathcal{S}$ is the set of desired graph matrices, e.g., adjacency matrices, combinatorial Laplacian matrices, etc. To do so, we can employ methods existing in the GSP literature that, given the graph matrix eigenbasis, retrieve a sparse matrix representation of the graph~\cite{segarra2017network,coutino2018sparsest}.

\textbf{One-Shot State Graph Estimation.} In alternative to the previous two cases, we can estimate the network topology related to the states by avoiding the computation of $\A_T$ explicitly. That is, after obtaining an estimate of $\C_T$, and hence $\bm T$, we can notice that system~\eqref{eq.systAt} can be modified to include the graph matrix, i.e.,
\begin{equation}\label{eq.systL}
	\bm U_{R,l}\bm T^{-1}\L_i\bm A = \U_{R,r}\bm T^{-1}\L_i,
\end{equation}
where $\U_{R,l}$ and $\U_{R,r}$ are the left and right matrices associated with $\U_R$ in~\eqref{eq.systAt}. Notice that in~\eqref{eq.systL}, we not only exploit the shift invariance in the $\U_R$ matrix but also the fact that $\L_i$ and $\A$ commute. We can check that this relation holds by recalling that [cf.~\eqref{eq.Ur}]
\begin{equation}
	\bm U_{R,l}\bm T^{-1}\L_i\A= \begin{bmatrix}
	\C\L_i\A\\ \C\A\L_i\A\\ \vdots \\ \C\bm A^{s-2}\L_i\A
	\end{bmatrix}=\begin{bmatrix}
	\C\A\L_i\\ \C\A^2\L_i\\ \vdots \\ \C\bm A^{s-1}\L_i
	\end{bmatrix} = \U_{R,r}\bm T^{-1}\L_i.
\end{equation}

As a result, we can pose the following optimization problem
\begin{equation}\label{eq.optOneShot}
    \begin{array}{ll}
    \underset{\L_i\in\mathcal{S},\bm M\in\mathcal{M}}{\min} & \Vert \U_{R,l}\bm T^{-1}\bm M - \U_{R,r}\bm T^{-1}\L_i \Vert_{\rm F}^2 + \mu\Vert \L_i\Vert_{1}
    \end{array},
\end{equation}
where we have defined $\bm M := \L_i\A$ to convexify the problem. Here, $\mu$ is a regularization parameter controlling the sparsity of $\L_i$ and the optimization is carried out over the set of desired graph matrices, $\mathcal{S}$, (as in~\eqref{eq.Lnorm0}) and $\mathcal{M}$ is a convex set of matrices meeting conditions derived by the matrix representation of the graph, e.g., if $\L_i$ is restricted to a combinatorial Laplacian then $\bm 1^T\bm M = \bm 0^T$ must hold. Alternatively, we could solve for $\L_i$ and $\bm A$ by means of alternative minimization~\cite{grippo2000convergence}.

Although in principle this approach requires knowledge of $\bm T$, in many instances it is possible to find a graph matrix associated with the transformed system, i.e., a graph associated with the system $\{\A_T,\B_T,\C_T,\D\}$, as the shift invariance property is oblivious to this ambiguity.
\setcounter{figure}{0}
\begin{figure}
    \begin{subfigure}{0.25\textwidth}
        \centering
        \includegraphics[width = 0.95\textwidth]{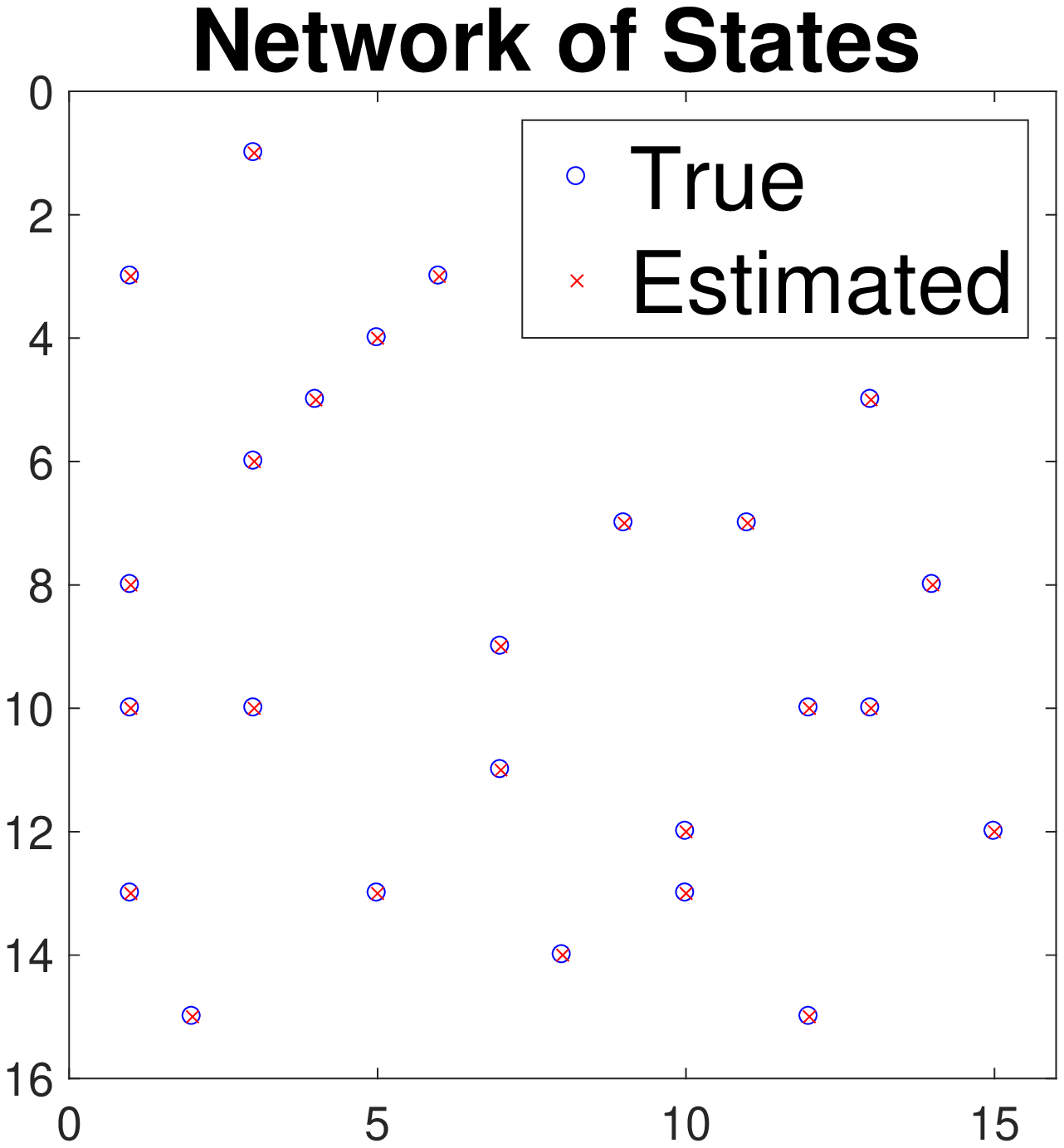}
        \includegraphics[width = \textwidth]{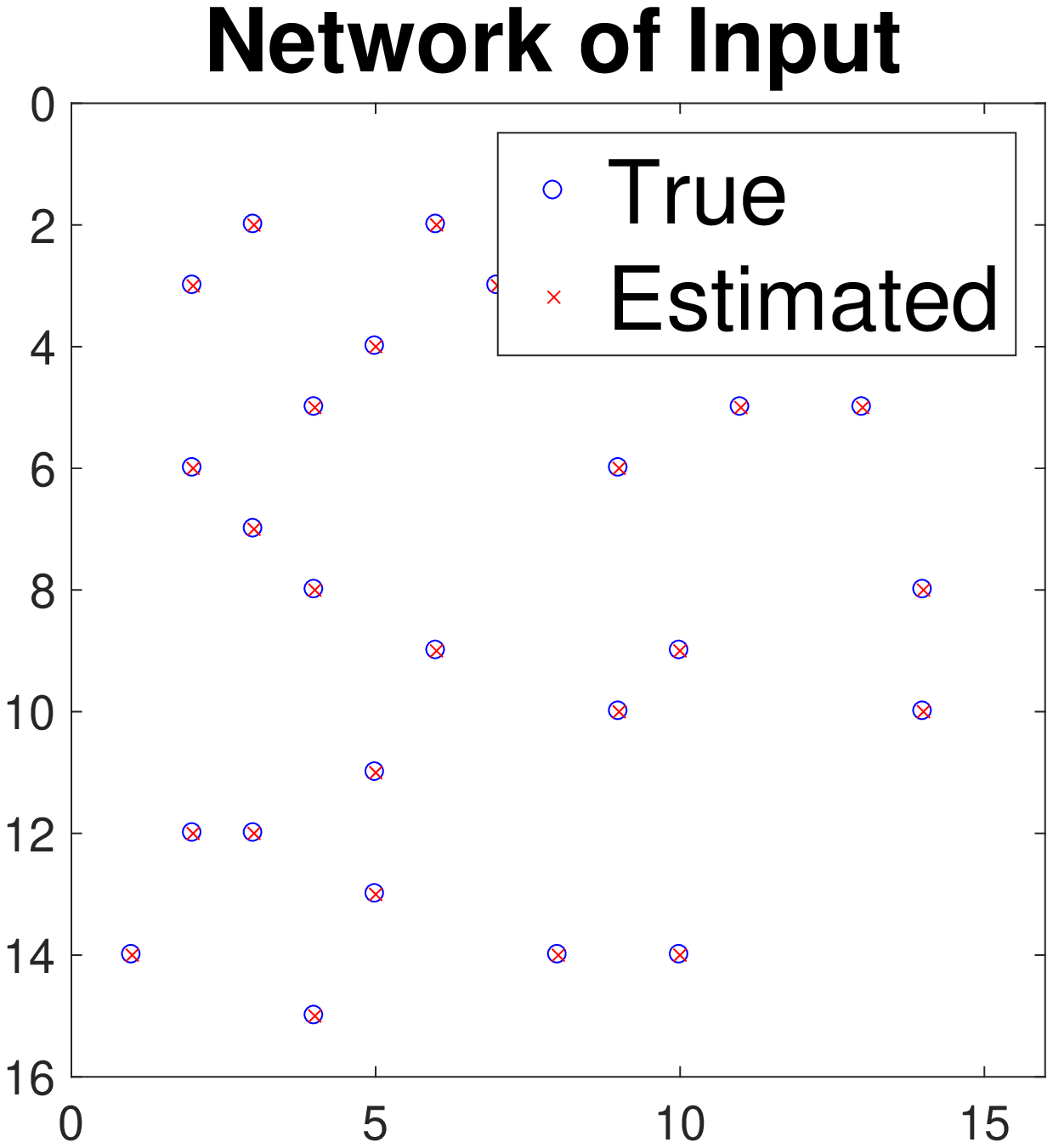}
        \caption{}
        \label{fig.gg2}
    \end{subfigure}%
    \begin{subfigure}{0.25\textwidth}
        \centering
        \includegraphics[width = \textwidth]{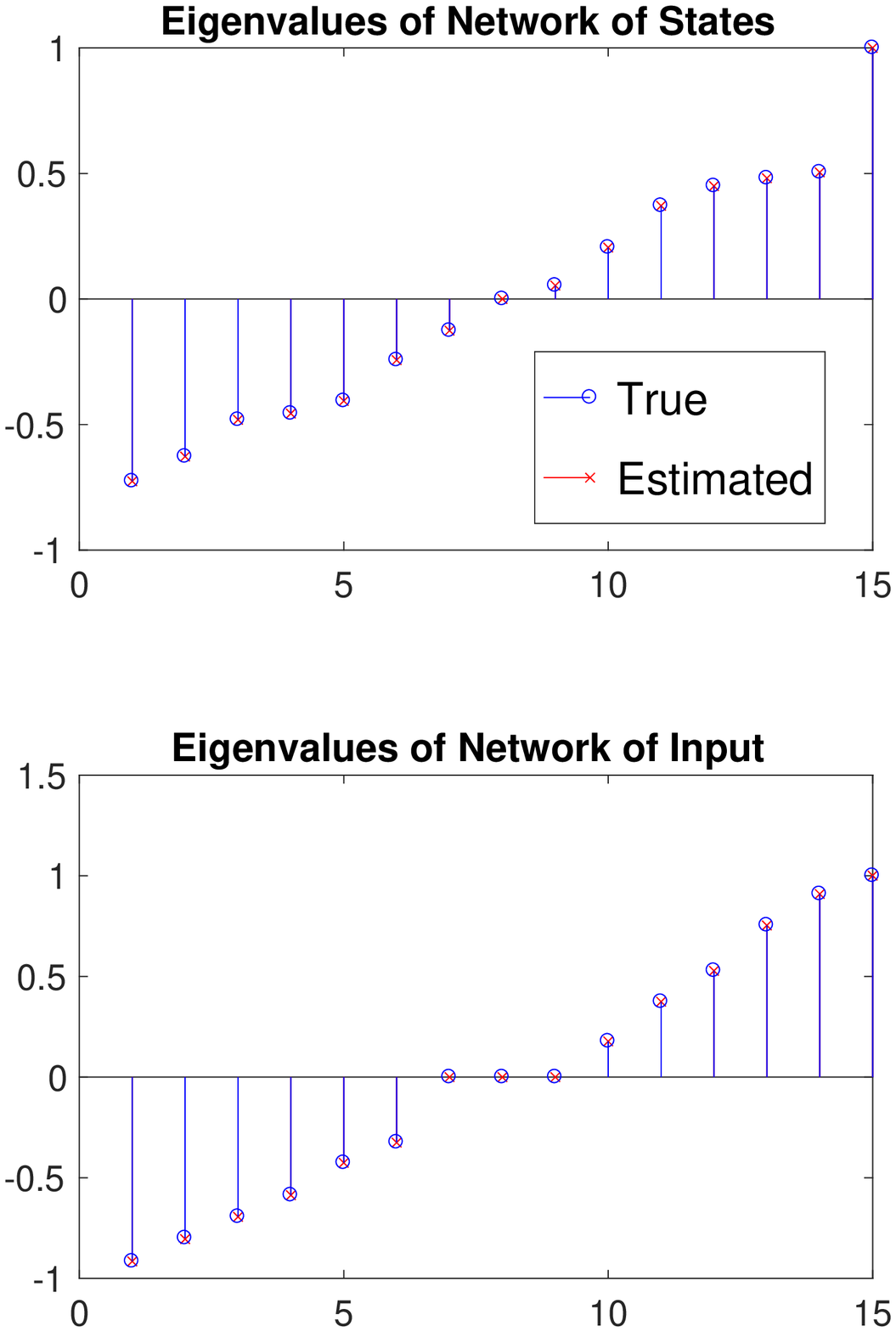}
        \caption{}
        \label{fig.gg1}
    \end{subfigure}
    \caption{Comparison of the true (blue circles) and estimated (red crosses) graphs for both the states and input. (a) Adjacency matrices for the states graph (top) and the input signal graph (bottom). (b) Eigenvalues for both states and input graph.}
    \label{fig:my_label}
    \vspace{-1mm}
\end{figure}
\vspace{-3ex}
\section{Numerical Examples} \label{sec.num}
To illustrate the performance of the proposed framework, we carry out a pair of experiments using synthetic and real data.

\textbf{Synthetic Example.} For this example, we consider a simple system where $\L_i \neq \L_j$ with $n=15$ nodes, $f_1^{\rm s}$ is a scaled diffusion map (i.e., $f_1^{\rm s}(x) = \alpha_ie^{-x\tau_i}$, also known as heat kernel), and $f_2^{\rm s}$ is the identity map, i.e., $f_2^{\rm s}(x) = x$. Here, it is assumed that all states are measured, i.e., $\bm C = \bm I$, and that there is a direct feedback from the input to the observations, i.e., $\bm D = \bm I$. As input, we considered a random piece-wise constant (during the sampling period) binary bipolar signal with $300$ samples each. The reconstruction of the topology using the proposed framework is shown in Fig.~\ref{fig:my_label}. In Fig.~\ref{fig.gg2}, the true and reconstructed adjacency matrices for the states and input are shown. As expected, when the data follows a practical model, the reconstruction of the matrices $\L_1$ and $\L_2$ is guaranteed to be exact. Here, since we have considered simple scalar mappings, we only perform root finding to retrieve the eigenvalues of the graph matrices. The eigenvalues comparison for both graphs is shown in Fig.~\ref{fig.gg1}.

\textbf{ETEX dataset.} We now consider data from the European Tracer Experiment (ETEX)~\cite{nodop1998field}. In this experiment a tracer was released into the atmospheric system and its evolution was sampled and stored from multiple stations in time. As it is unlikely that such kind of process has as many states as stations, we cluster the $168$ measuring stations in $25$ geographical regions and aggregate its measurements as a preprocessing step. This preprocessing is sustained by looking at the singular values of $\frac{1}{m}\bm Y_m$ in Fig.~\ref{fig.gg2}. In this figure, it is observed that most of the dynamics can be described with a system of order $5$, i.e., first knee in the plot. Here, we selected $25$ nodes as a trade off between complexity and graph interpretability (second knee). As the propagation of the tracer is considered to be a pure diffusion in an \emph{autonomous} system, i.e., matrices $\B$ and $\D$ equal zero, we employ the proposed one-shot state graph estimation method [cf.~\eqref{eq.optOneShot}] to retrieve the underlying network structure. In this case, it is also assumed that the observations are the states of the system, i.e., $\bm C = \bm I$. The estimated graph is shown in Fig.~\ref{fig.g2}. Here, the size of the circle representing a vertex is proportional to the degree of the node. From the estimated graph, we can observe that the region of Berlin presents the highest degree which is consistent with the concentration results in~\cite{thanou2017learning}. Further, the strong connectivity along the France--Germany region correlates with the spreading pattern of the agent. Despite that this graph has fewer nodes than the one obtained in~\cite{thanou2017learning} (see Fig.~\ref{fig.g3}), the estimated graph presents a better visual interpretability and exhibits a similar edge behaviour.
\vspace{-2ex}
\section{Conclusion}

In this paper, we have introduced a general framework for graph topology learning using state space-models and subspace techniques. Specifically, we have shown that it is possible to retrieve the matrix representation of the involved graphs from the system matrices by different means. In the particular case of the graph related to the states, we presented a one-shot method for topology identification that does not require the explicit computation of the system matrix. Numerical experiments for both synthetic and real data have demonstrated the applicability of the proposed method and its capabilities to recover the topology of the underlying graph from data.

{\footnotesize
\bibliographystyle{IEEEbib}
\bibliography{dissertation}
}
\end{document}